\documentclass[12pt]{iopart}
\usepackage{graphicx}% Include figure files 
\begin{document}

\title[Suppressed Magnetism at the Surfaces and Interfaces of Manganites]{Suppressed Magnetization at the Surfaces and Interfaces of Ferromagnetic Metallic Manganites}

\author{J.W. Freeland$^1$,  J. J. Kavich$^{1,2}$, K.E. Gray$^3$, L. Ozyuzer$^3$, H. Zheng$^3$, J.F. Mitchell$^3$, M. P. Warusawithana$^4$, P. Ryan$^2$ X. Zhai$^4$, R. H. Kodama$^2$, and J. N. Eckstein$^4$}
\address{$^1$Advanced Photon Source, Argonne National Laboratory, Argonne, IL 60439}
\address{$^2$Department of Physics, University of Illinois at Chicago, Chicago, IL 60607}
\address{$^3$Materials Science Division, Argonne National Laboratory, Argonne, IL 60439}
\address{$^4$Department of Physics, University of Illinois at Urbana-Champaign, Urbana, IL 61801}
\ead{freeland@anl.gov}
\begin{abstract}
What happens to ferromagnetism at the surfaces and interfaces of manganites? With the competition between charge, spin, and orbital degrees of freedom, it is not surprising that the surface behavior may be profoundly different than that of the bulk. Using a powerful combination of two surface probes, tunneling and polarized x-ray interactions, this paper reviews our work on the nature of the electronic and magnetic states at manganite surfaces and interfaces. The general observation is that ferromagnetism is not the lowest energy state at the surface or interface, which results in a suppression or even loss of ferromagnetic order at the surface. Two cases will be discussed ranging from the surface of the quasi-2D bilayer manganite (La$_{2-2x}$Sr$_{1+2x}$Mn$_2$O$_7$) to the 3D Perovskite (La$_{2/3}$Sr$_{1/3}$MnO$_3$)/SrTiO$_3$ interface. For the bilayer manganite, that is, ferromagnetic and conducting in the bulk, these probes present clear evidence for an intrinsic insulating non-ferromagnetic surface layer atop adjacent subsurface layers that display the full bulk  magnetization. This abrupt intrinsic magnetic interface is attributed to the weak inter-bilayer coupling native to these quasi-two-dimensional materials. This is in marked contrast to the non-layered manganite system (La$_{2/3}$Sr$_{1/3}$MnO$_3$/SrTiO$_3$), whose magnetization near the interface  is less than half the bulk value at low temperatures and decreases with increasing temperature at a faster rate than the bulk. 
\end{abstract}

\maketitle

\section{Introduction}
One of current problems of intense interest to the condensed matter community is strongly correlated electron systems. These systems contain a variety of competing strong interactions, which create a subtle balance to define the lowest energy state.  The broken symmetry at a surface or interface can then drastically upset the subtle balance of competing energies in a strongly correlated electron system and lead to significant deviations from the bulk properties. Manganites have been extensively studied for many years because of the extensive range of dynamic electronic, magnetic, and structural properties that arise from these strong electron correlations  \cite{mbsreview}, e.g., colossal magneto-resistive behavior associated with both the ferromagnetic (FM) Curie temperature, Tc, and the metal-insulator transition  \cite{cmr1,cmr2}. 
\begin{figure}[h]
\begin{tabular}{cc}
	\includegraphics[scale=.4]{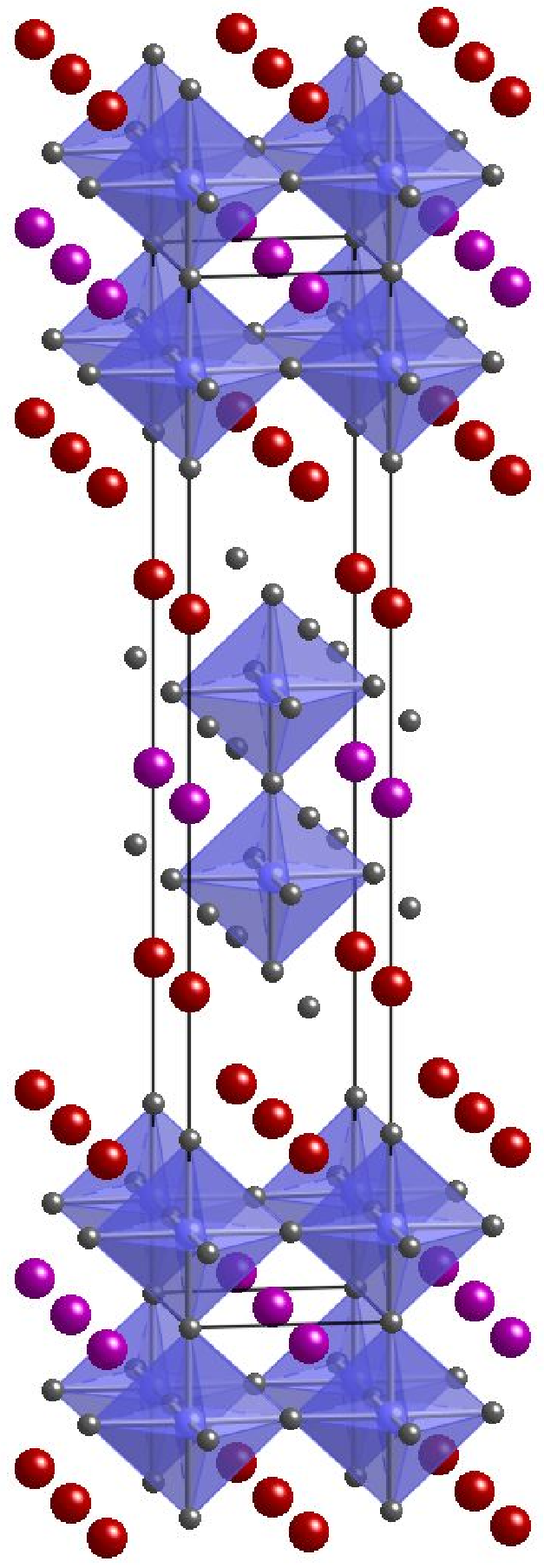} &
	\includegraphics[scale=.4]{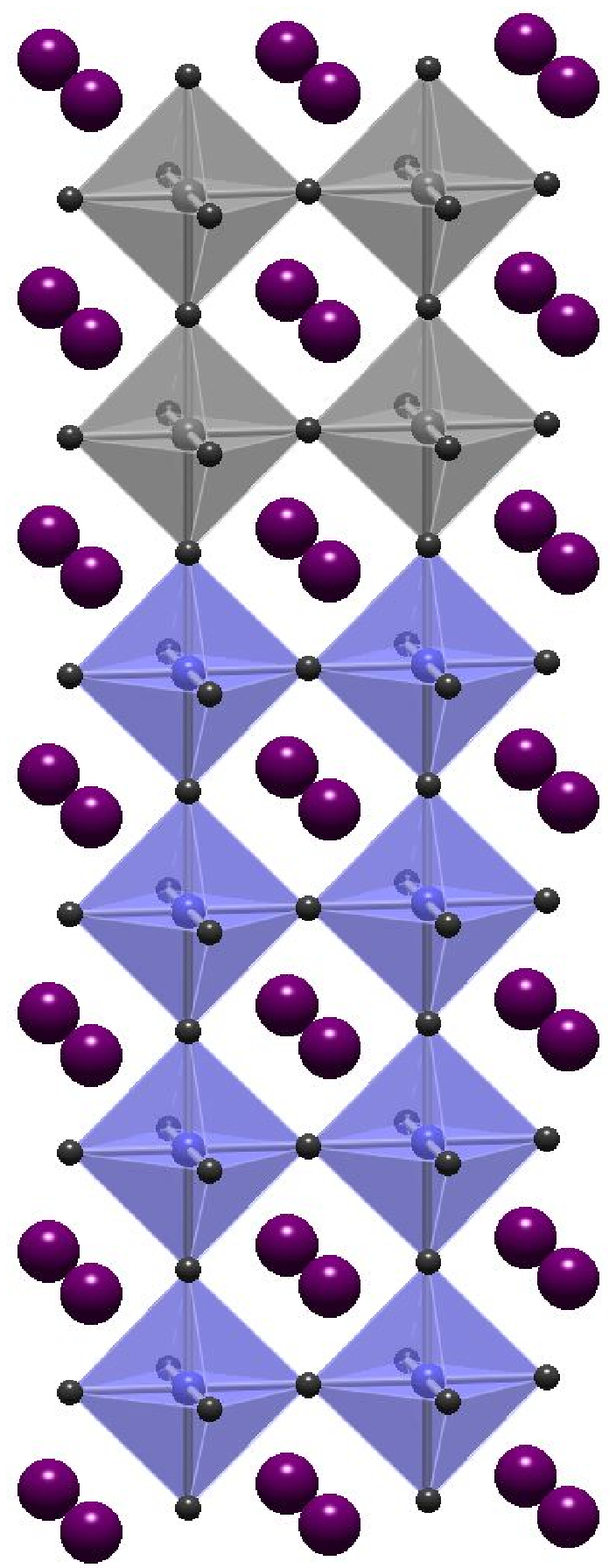} \\
	(a) La$_{2-2x}$Sr$_{1+2x}$Mn$_2$O$_7$&
	(b) La$_{2/3}$Sr$_{1/3}$MnO$_3$/SrTiO$_3$ \\
\end{tabular}
\caption{(a) Crystal structure for bilayer manganite  and (b) La$_{1-x}$Sr$_x$MnO$_3$/SrTiO$_3$ interface.}
\protect\label{crystal} 
\end{figure}

The case of ferromagnetic (FM) manganites has attracted significant attention due to the high spin polarization at the Fermi level. The hole doped La$_{1-x}$Sr$_x$MnO$_3$ (LSMO) with x = 0.3 is FM and nearly half metallic at low temperatures. Experiments indicate that the spin polarization of these materials is at least 0.83 \cite{ref2,ref3,ref4,ref5,ref6} (compared with ~0.45 for FM Fe, Ni, Co \cite{ref3,ref4}) with some evidence for it being completely spin polarized \cite{mbprl,ref7,ref8,ref9}. 
At finite temperature, the story changes significantly. Several groups have investigated magnetic tunnel junctions (MTJs) using manganite films as electrodes and have observed a significant reduction in tunneling magneto-resistance (TMR) between 200K and 280K \cite{ref5,ref8,ref10,ref11,ref11_1} even though the bulk displays a sizable magnetization. Spin resolved photoemission studies indicate that a strong possibility for this loss of TMR is a degraded magnetization at the surface \cite{ref12}. 

For the case of free surfaces, experiments have attributed this to the possibility of cation segregation, which changes the electronic structure in the surface region \cite{ref13,ref14}. On the theoretical side, several studies have examined the problem in detail. For the case of a free surface, predictions have been made for modified magnetic interactions due to surface orbitals \cite{theory1}, the influence of strain on the phase diagram \cite{theory2}, and a picture of how reduced hopping at the surface leads to an insulating surface layer \cite{theory3}. Recently, the case of how the electronic and magnetic structure change at various interfaces has also been examined \cite{theory4}. 

Here we present our recent work on measurements of the magnetization profile near manganite surfaces and interfaces. The general observation is that ferromagnetism is not the lowest energy state at the surface or interface, which results in a suppression or even loss of ferromagnetic order at the surface. In this paper,  two cases will be discussed ranging from the surface of the quasi-2D bilayer manganite (La$_{2-2x}$Sr$_{1+2x}$Mn$_2$O$_7$) \cite{natmat} to the 3D Perovskite (La$_{2/3}$Sr$_{1/3}$MnO$_3$)/SrTiO$_3$ interface \cite{jjklsmo} (see Fig.\ \ref{crystal}).

\section{Experiment}

 \begin{figure}[h]
\includegraphics[scale=.4]{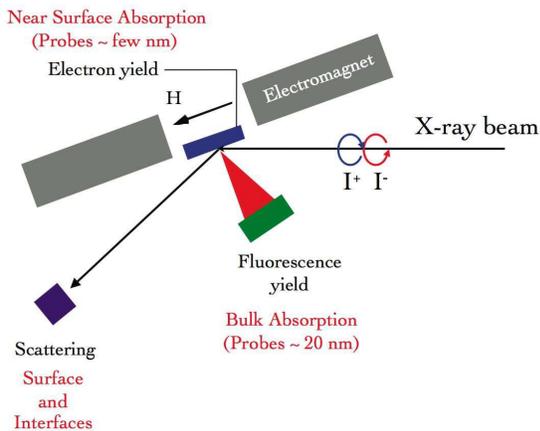}
\caption{Diagram showing the experimental configuration for simultaneous measurement of x-ray absorption (near surface and bulk sensitive) and resonant scattering while switching the polarization between left-circular-polarization (LCP) and right-circular-polarization (RCP). }
\protect\label{expt} 
\end{figure}
\begin{figure}[h]
\includegraphics[scale=.4]{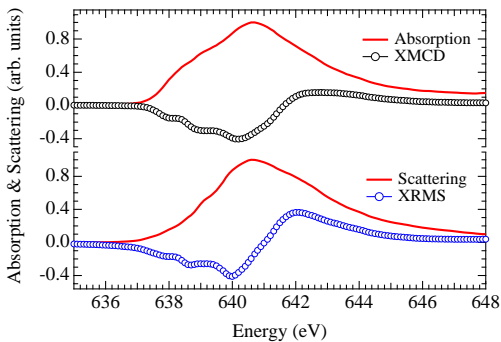}
\caption{Comparison of near surface x-ray absorption (top) and scattering (bottom). The magnetic signals are shown as XMCD and XRMS, respectively. These data were taken at 30K in an applied field of 0.05 T for the bilayer maganite for an incident angle of $\theta$ = 11 deg.}
\protect\label{xmcdxrms} 
\end{figure}

 
To characterize the interface magnetism, polarized x-ray techniques were used at beamline 4-ID-C of the Advanced Photon Source \cite{blrsi}. The experimental configuration (see Fig.\ \ref{expt}) allows for simultaneous measurement of x-ray absorption and scattering while switching the polarization between left-circular-polarization (LCP) and right-circular-polarization (RCP).   The sum (I$^+$+I$^-$) of these signals provides information on the electronic environment of the Mn 3d electrons while magnetic information is contained in the difference (I$^+$-I$^-$), which in absorption and scattering are referred to respectively as as x-ray magnetic circular dichroism (XMCD) \cite{xmcd} and x-ray resonant magnetic scattering(XRMS) \cite{xrms,jbk} (see Fig.\ \ref{xmcdxrms}).  The absorption always provides a measurement that is an average over the probing depth of the measurement mode. Measurements of absorption by monitoring the photocurrent due to escaping electrons, referred to as  total electron yield (TEY), probe very close to the surface or interface due to the short probing depth (1-2 nm), although some studies indicate larger probing depths for oxides \cite{feoxidetey}. To probe the bulk, total flourescence yield (TFY) is used due to the long probing depth of the escaping x-rays ($\sim$100 nm). However, both of these measurements present only an average of the properties over either the near surface (TEY) and bulk (TFY). To understand the magnetization profile, we turn to resonant scattering probes. \cite{drlprb}
By tuning to a Mn L edge, where the chemical and magnetic scattering factors from the Mn 3d are of a similar magnitude \cite{xrms}, scattering from a FM layer deeper down will then interfere with scattering from the chemical interface and result in interference wtih a corresponding change in the XRMS lineshape. This interference can be measured via angle dependent XRMS and can determine the thickness of the non-magnetic region near the surface or interface.   This will be discussed in more detail in Section \ref{xrms-sec}.

The La$_{2-2x}$Sr$_{1+2x}$Mn$_2$O$_7$ single crystals were prepared by the floating zone method  \cite{jfm1,jfm2} between x=0.36 and x=0.4.  Over this range we found no substantive differences in any measured properties with x except for T$_C$, which has a known doping dependence \cite{jfm2}. For surface studies, the natural cleavage planes of the layered manganites, La$_{2-2x}$Sr$_{1+2x}$Mn$_2$O$_7$, provide a clear advantage (see Fig.\ \ref{crystal}(a)).  In addition, their bulk properties have been studied extensively by x-ray and neutron scattering, transport, and magnetization \cite{jfm1,jfm2}.  In particular, all are double-exchange ferromagnets that exhibit symbiotic ferromagnetism (FM) and metallic conductivity below T$_C$ of ~120 K.  Crystals were cleaved just prior to mounting in the respective apparatuses, and both atomic-force microscopy and soft-x-ray rocking curves showed large, flat terraces.

Due to the lack of a cleavage plane for La$_{2/3}$Sr$_{1/3}$MnO$_3$, the samples used in our experiment were grown using ozone-assisted atomic layer-by-layer molecular beam epitaxy (ALL-MBE) \cite{mbe} on a (100) oriented SrTiO$_3$ (STO) substrate. The atomic beam fluxes were calibrated to better than 0.1$\%$ 
accuracy using Rutherford backscattering spectroscopy and x-ray film thickness oscillations on separate films. The growth was carried out at a substrate temperature of 680$^\circ$C and an ozone pressure of 2x10$^{-6}$ torr using flux matched co-deposition. Throughout the LSMO growth, RHEED indicated intensity oscillations of the specular reflection characteristic of a 2-dimensional growth mode. X-ray diffraction studies on similar films have shown that the LSMO grows pseudomorphic to the STO substrate. Immediately after the growth of the La$_{2/3}$Sr$_{1/3}$MnO$_3$) (LSMO) layer, one film was capped with 2 unit cells (~0.8 nm) of STO (LSMO/STO interface) as shown in Fig.\ \ref{crystal}(b).  In a previous study of such Òsingle crystalÓ MTJs comprised of a thin epitaxial CaTiO$_3$ or SrTiO$_3$ tunnel barrier sandwiched between two La$_{2/3}$Sr$_{1/3}$MnO$_3$ electrodes \cite{ref3}, large values of TMR exceeding 450$\%$ at 15K were obtained. The structures used to measure the interface magnetization were prepared identically to those MTJs.
 
 \section{X-ray Resonant Magnetic Scattering}
 \label{xrms-sec}
  
\begin{figure}[h]
  	\begin{tabular}{cc}
		\includegraphics[scale=.4]{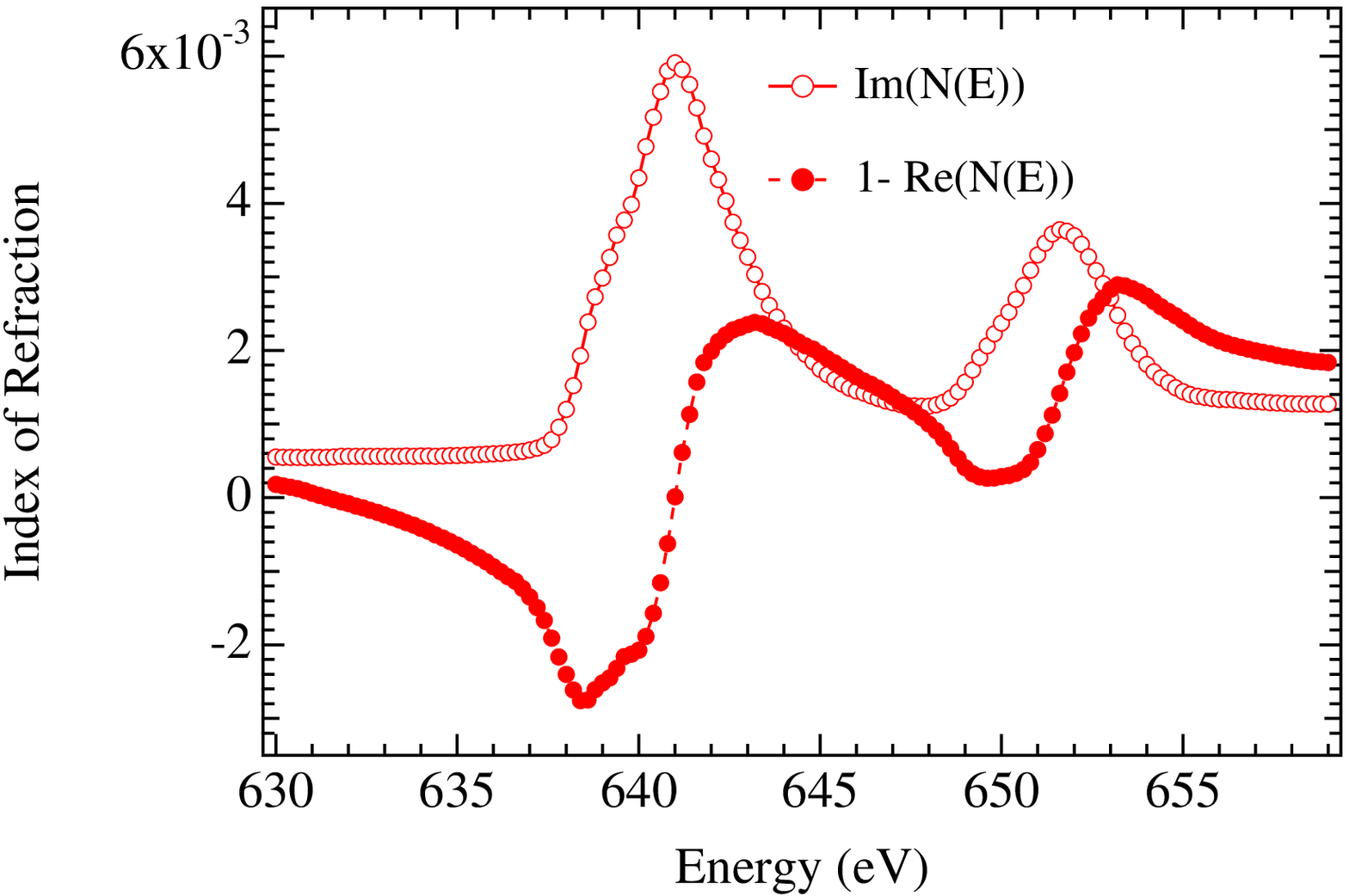} &
		\includegraphics[scale=.4]{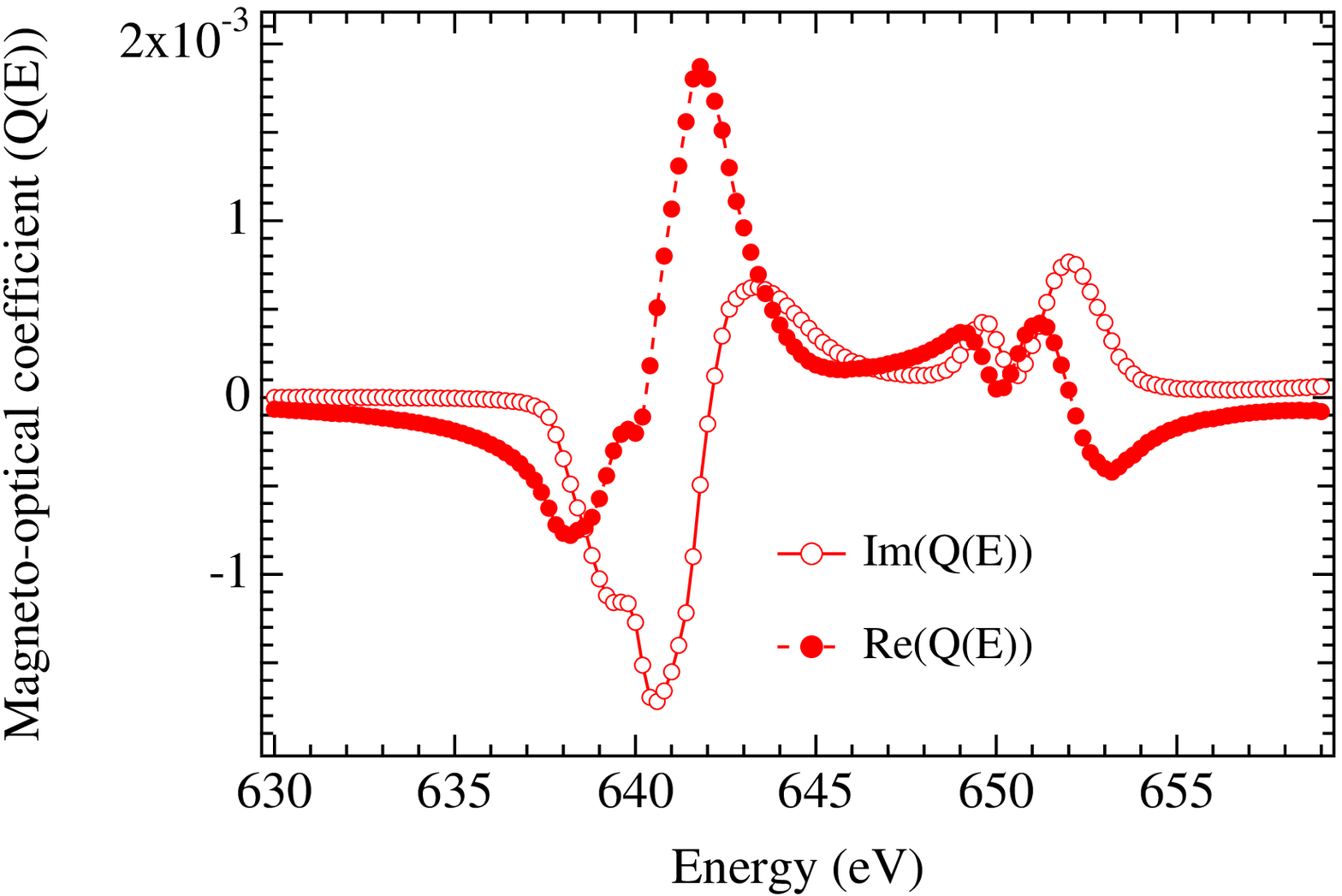} \\
		(a) &
		(b) \\
	\end{tabular}
\caption{The real and imaginary parts for the MnO$_2$ dielectric tensor near the Mn L edge. (a) Index of refraction $N(E)$ and(b) magneto-optical coefficient $Q(E)$.}
\protect\label{nqvsE} 
\end{figure}

Understanding of  the ability of X-ray Resonant Magnetic Scattering (XRMS) spectra to measure magnetic profiles, requires a discussion on how the scattering process works. XRMS results from the fact that the energy dependent dielectric tensor of a magnetic material contains off-diagonal elements that have a strong energy dependence near an absorption resonance \cite{xrms,jbk}. Due to the long wavelength of the soft x-rays (1-2 nm) with respect to inter-atomic spacings, the sample can be described as a continuous medium  and the scattering can be described by the same formalism as the magneto-optical Kerr effect \cite{moke1,moke2}. The starting point to understanding  XRMS rests on the resonant behavior of the dielectric tensor. For the geometry used here (referred to as longitudinal), the magnetic moment lies in the plane of  the film. In this case the dielectric tensor contains the following elements
\begin{equation}
\epsilon (E)=N(E)^2\left( {\matrix{{1}&{0}&{-iQ(E)}\cr
{0}&{1}&{0}\cr
{-iQ(E)}&{0}&{1}\cr
}} \right),
\vspace{0.1in}
\end{equation}
where N is the energy dependent index of refraction, and Q is the magneto-optical coefficient. To determine the proper dielectric tensor requires measurement of $N(E)$ and $Q(E)$. Using the fact that $N(E)$ and $Q(E)$ are complex and that the imaginary parts relate to the absorption and XMCD, respectively, we can reconstruct the real parts using a Kramers-Kronig transformation (see Fig.\ \ref{nqvsE}). This is done quantitatively by scaling the data to the tabulated data for the nonresonant scattering factors \cite{henke}. One point to note here is that at this stage we have assumed that the material is electronically isotropic which is not always true in the case of complex oxides.

To construct the scattering from the corresponding dielectric tensor, we need to determine the corresponding reflectivity of the sample at a given angle $\theta$ and energy E represented as
\begin{equation}
R(\theta,E)=\left( {\matrix{{r_{ss}(\theta,E)}&{r_{sp}(\theta,E)}\cr
{r_{ps}(\theta,E)}&{r_{pp}(\theta,E)}\cr
}} \right)
\end{equation}
where $r_{ss}$ and $r_{pp}$ are the reflection coefficients for s and p polarized light and $r_{sp}$ and $r_{ps}$ are magnetic reflectivity terms that mix the s and p polarization states. With an incoming circularly polarized photon described as
\begin{equation}
E_{In}^{\pm}=A\left( {\matrix{1\cr
{\mp i}\cr
}} \right),
\end{equation}
the helicity dependent scattered intensity is then
\begin{equation}
I^{\pm} = \left| {R(\theta,E) \cdot E_{In}^{\pm}} \right|^2
\end{equation}
From this the sum and difference spectra
are then determined as
\begin{equation}
\left(I^++I^-\right)=A^{2}\left[\left| {r_{ss}} \right|^2+ \left|
{r_{pp}} \right|^2+\ldots\right] and
\protect\label{chemscatt}
\end{equation}
\begin{equation}
\left(I^+-I^-\right)=-4A^{2}
Im\left[r_{ss}^{*}r_{sp}+r_{pp}r_{ps}^*\right],
\protect\label{magscatt}
\end{equation}
where in the sum higher order magnetic terms are small and can be
ignored. Equations\ \ref{chemscatt} and
\ref{magscatt} then show clearly that ($I^++I^-$) is purely chemical while
($I^+-I^-$) contains both chemical ($r_{ss}$ and $r_{pp}$) and magnetic
scattering
contributions ($r_{sp}$ and $r_{ps}$), which directly indicates that it is
not purely magnetic
in origin.

For the case of a magnetic surface the analytical form for the elements composing R are known \cite{thinfilm,kerrapl}. However, due to the strong multiple scattering in the soft x-ray regime, analytical forms are not simple to construct even for a single layer film \cite{thinfilm}. In order to simulate the XRMS data, we following the MOKE formalism using boundary matrices as outlined by Zak et al. \cite{moke1,moke2}. By representing the sample in a slab structure, we can then compute R($\theta$,E) for a given magnetic structure and use the equations above to construct the scattered intensity to compare with the experimental data.

\begin{figure}[h]
  	\begin{tabular}{cc}
		\includegraphics[scale=.4]{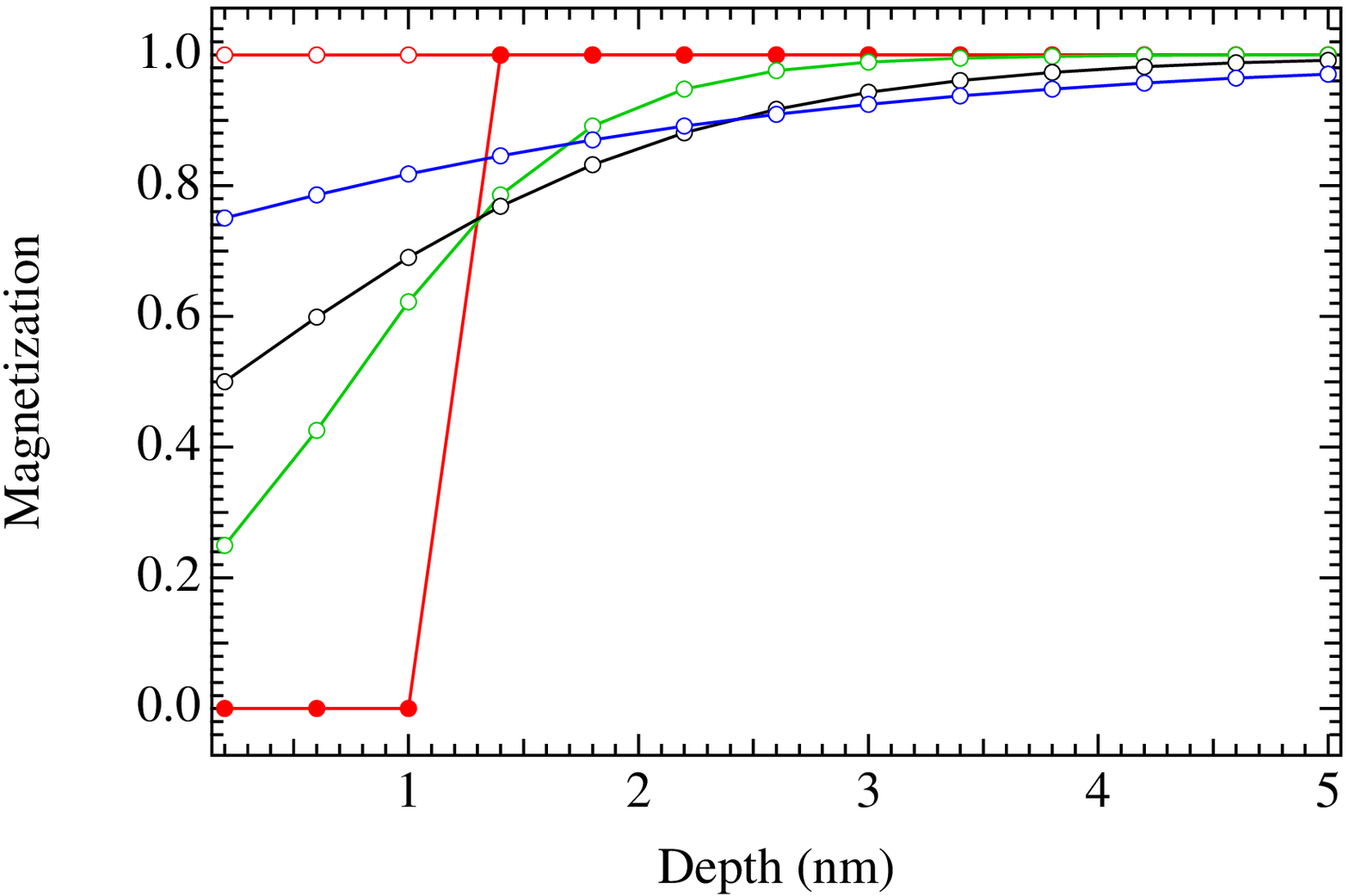} &
		\includegraphics[scale=.4]{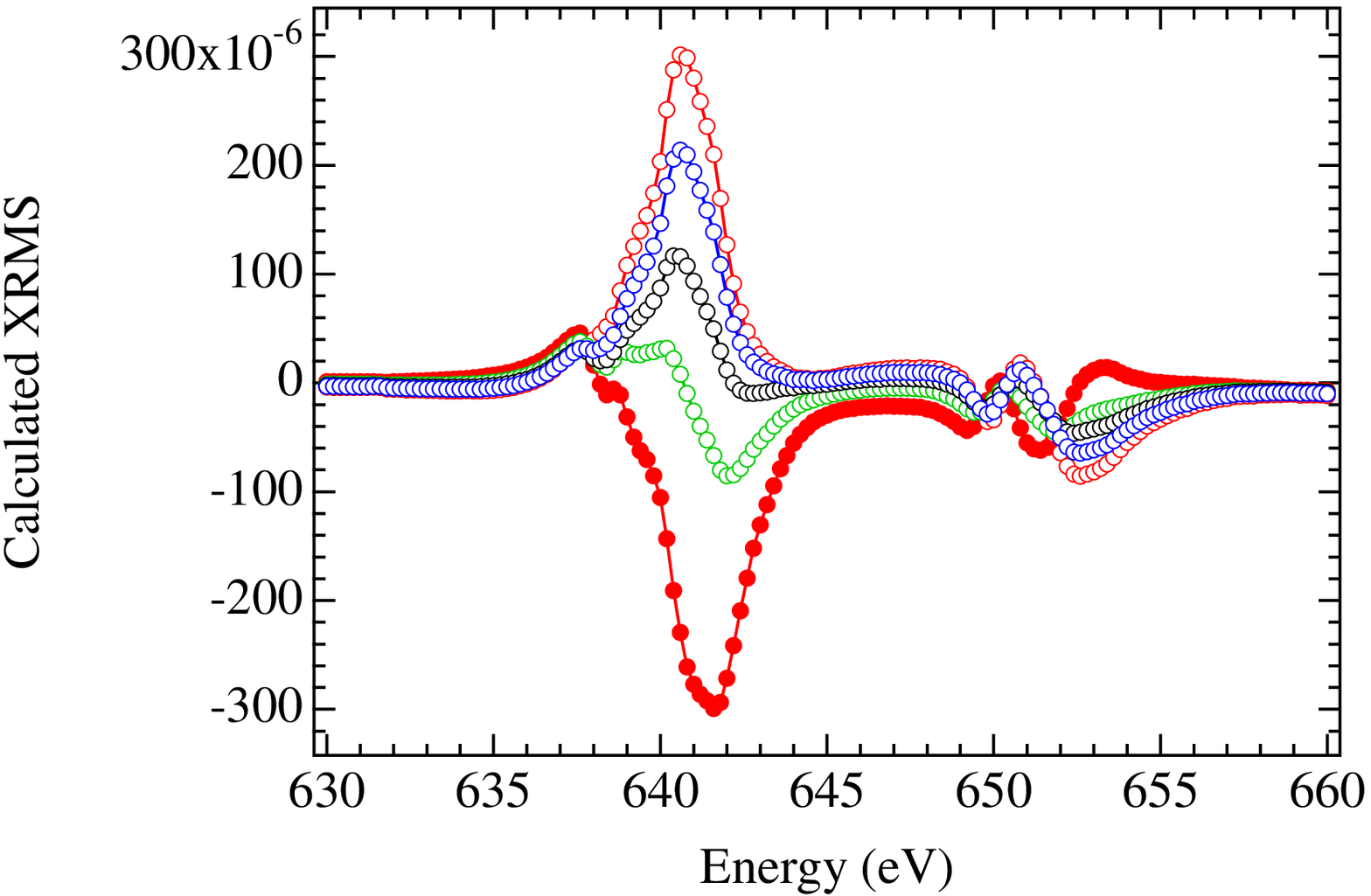} \\
		(a) &
		(b) \\
	\end{tabular}
\caption{(a) Simulated magnetic profiles near the surface a 40 nm LSMO film and (b) the corresponding calculated XRMS for an incident $\theta$ = 11 degrees. Note the large change in lineshape due to the changing magnetic profile.}
\protect\label{calcxrms} 
\end{figure}

Even though the incident soft x-ray wavelength is $\sim$2 nm, the scattered intensity is quite sensitive to the magnetic profile even within 1nm of the surface (see Fig.\ \ref{calcxrms}). This is due to the fact that at grazing incidence, the photon still traverses several wavelengths within the sample. It is this sensitivity that we will make use of in the follow sections to study the suppression of ferromagnetism in the manganites.

 \section{Quasi-2D Layered Manganite (La$_{2-2x}$Sr$_{1+2x}$Mn$_2$O$_7$)}
 
In the bulk at doping levels near x=0.36, the bilayer manganite resides in a ferromagnetic metallic groundstate \cite{jfm1,jfm2}. Due to the natural cleavage plane, the double-layered manganite offers an excellent template on which one can study any influence of the surface on the ground-state. The cleavage results in a high-quality surface with terraces of the order of 1 $\mu$m size with the lowest-order defect due to steps of N/2 unit cells (i.e., one bilayer) \cite{renner}. After creating this template, we can utilize a combination of probes to assess the nature of the electronic and magnetic state at the surface, which as will be seen below is quite distinct from the bulk state \cite{natmat}. 

\begin{figure}[h]
\includegraphics[scale=.4]{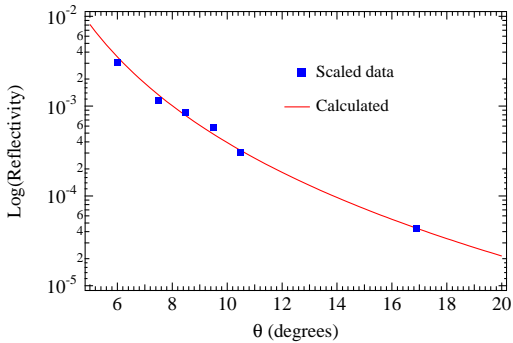}
\caption{Comparison of calculated and measured reflectivity (I$^+$ + I$^-$) at the Mn L$_3$ resonance (641 eV) as a function of incident angle. The intensity drops monotonically as expected for a surface following q$_z^{-4}$ with q$_z = 4\pi sin (\theta)/\lambda$}
\protect\label{scattvsth} 
\end{figure}
For the case of a surface, the average (I$^+$ + I$^-$) scattered intensity is expected to decay monotonically  with a power of  q$_z^{-4}$. This is precisely what is seen when we plot the scattering intensity when tuned to the Mn L$_3$ edge (641 eV) (see Fig.\ \ref{scattvsth}). To model scattering from the layered manganite, we utilized the approach described in Section\ \ref{xrms-sec}. The crystal structure shown in Figure 1 was converted to a continuous medium with the 2 nm unit cell represented by [MnO$_2$(0.5nm)/LaSrO(0.5 nm)/MnO$_2$(0.5 nm)/LaSrO(0.5 nm)]. Even though the modeled structure is quite simple, it reproduces well the angular dependence of the (I$^+$ + I$^-$) signal (see Fig.\ \ref{scattvsth}). This implies that the {\it charge density} in the surface bilayer is similar to the bulk although there might be subtle differences not seen in the angle dependence.

\begin{figure}[h]
  	\begin{tabular}{cc}
		\includegraphics[scale=.4]{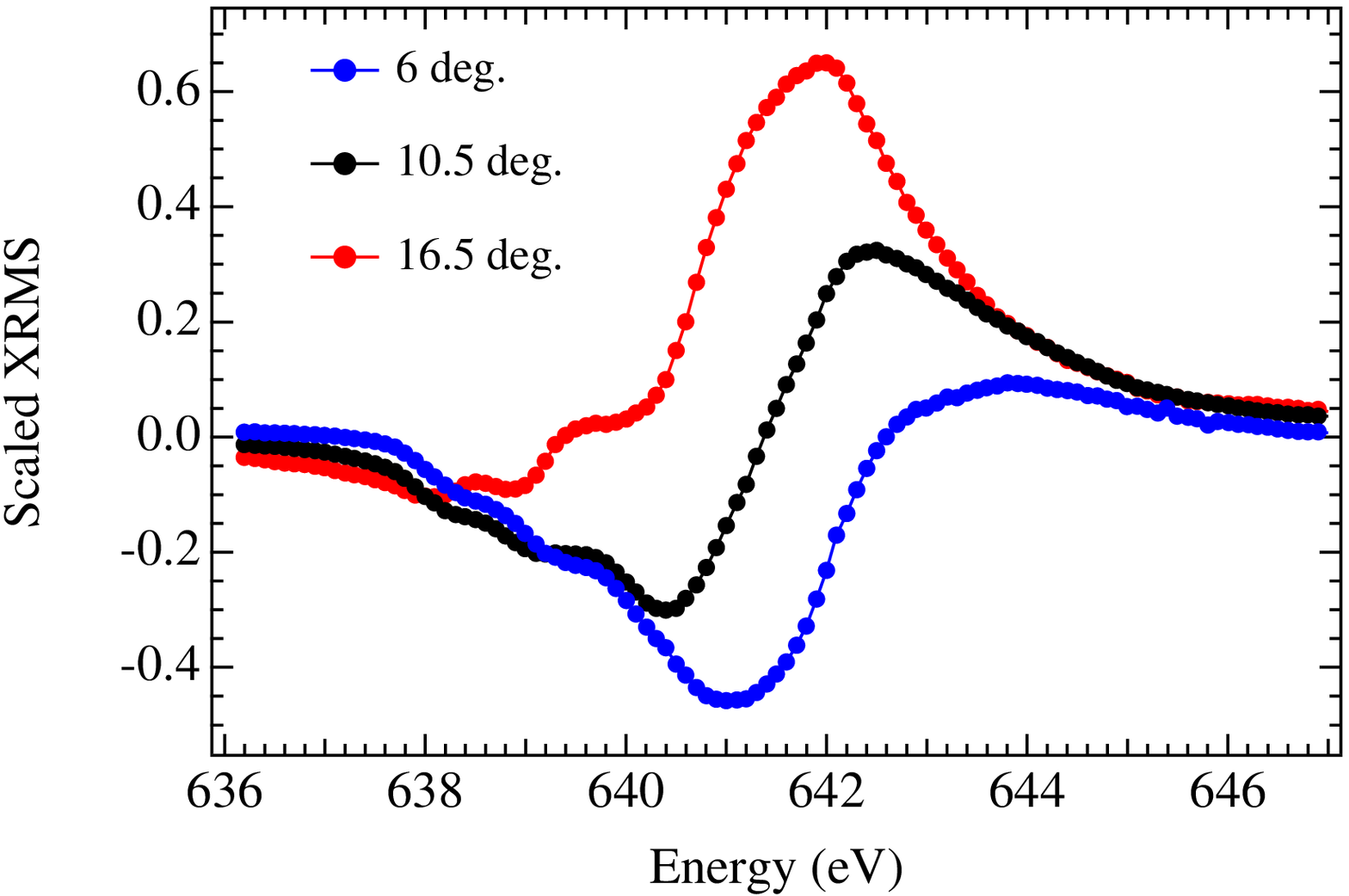} &
		\includegraphics[scale=.4]{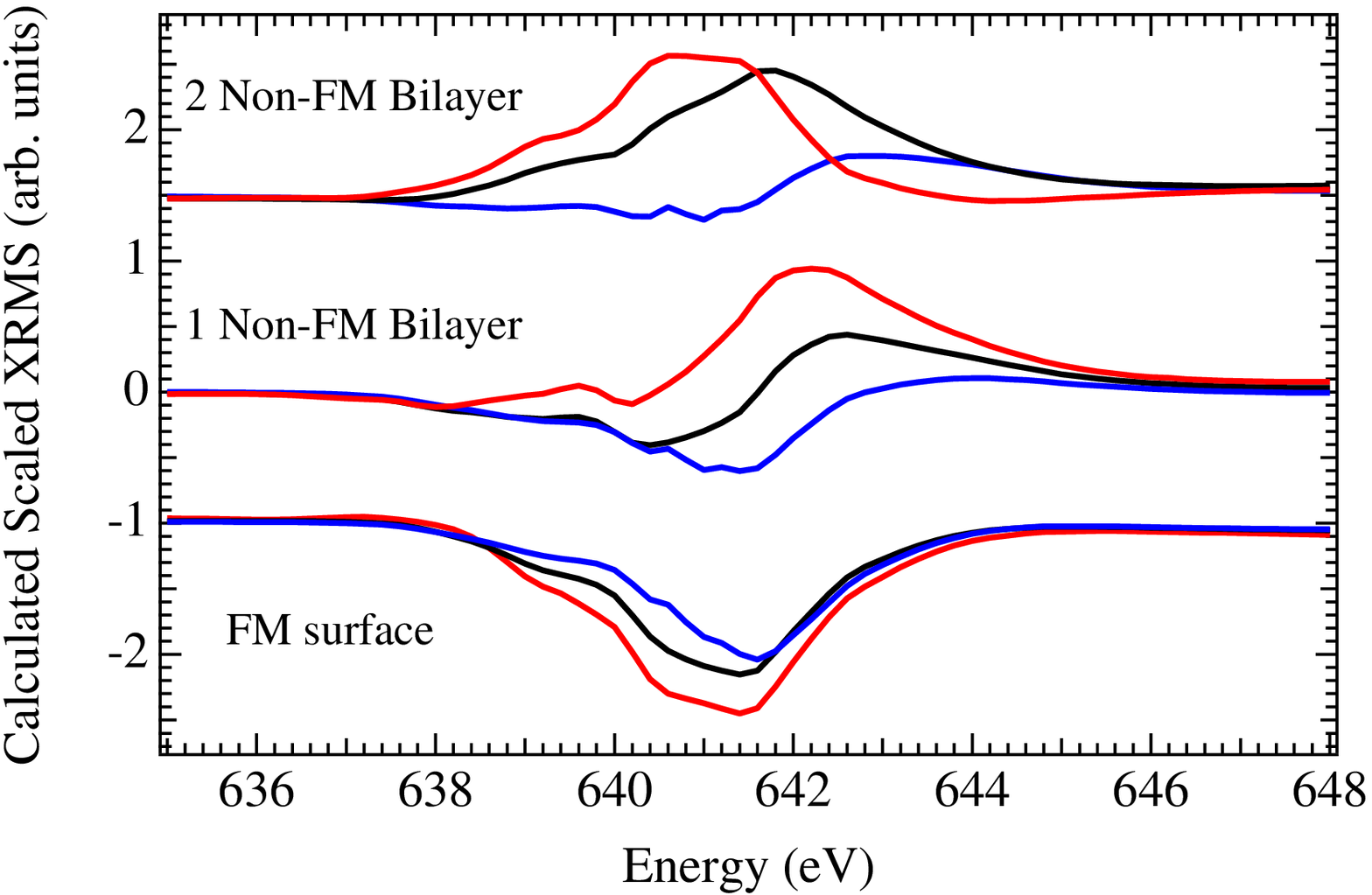} \\
		(a) &
		(b) \\
	\end{tabular}
\caption{(a) Scaled X-ray resonant magnetic scattering as a function of angle, which were all intensity normalized to an (I$^+$+I$^-$) of one. (b) Calculated energy and angle dependence of the XRMS for 3 cases: ferromagnetic (FM) surface, one non-ferromagnetic bilayer, and two non-ferromagnetic bilayers. Note, the colors in (b) indicate the angles as indicated in (a).}
\protect\label{scxrmsvsth} 
\end{figure}

If the magnetic profile differs from that of the charge, the presence of a non-ferromagnetic surface layer is readily observed because it gives rise to interference in the magnetic signal (XRMS) that is directly related to the thickness of the non-FM surface layer.  This interference between the chemical and magnetic interfaces manifests itself as a change in the sign of the XRMS data with incident angle, an effect that is clearly seen in Fig.\ \ref{scxrmsvsth}(a).  Given that the XRMS data change by an order of magnitude in total intensity, we have scaled the data in Fig.\ \ref{scxrmsvsth}(a) to an average scattering resonant peak height of one (see data in Fig.\ \ref{xmcdxrms}). For a coincident chemical and magnetic profile, the sign of the XRMS is strictly independent of the scattering angle. The strong dependence of calculated XRMS results provides a means to directly determine the magnetic profile through comparison with calculations for a variety of magnetic profiles (see Fig.\ \ref{scxrmsvsth}(b)).  Only the case of a single non-FM bilayer provides reasonable agreement with the data as shown in Fig.\ \ref{scxrmsvsth}(a).  The XRMS data thus reveal the presence of an extremely thin nonmagnetic Ônanoskin,Õ below which subsequent bilayers are magnetized.  

\begin{figure}[h]
\includegraphics[scale=.4]{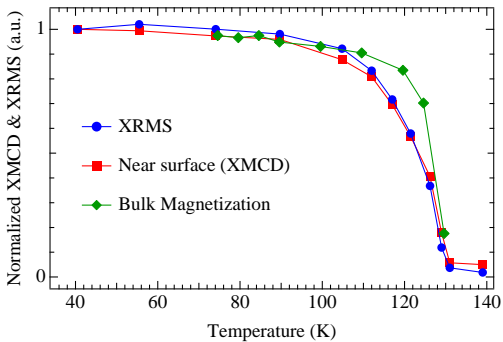}
\caption{Temperature dependence of the peak height of the XRMS and near surface XMCD compared with the bulk magnetization. }
\protect\label{xmcdxrmsvsT} 
\end{figure}

Having established that the topmost bilayer alone is non-ferromagnetic, we attribute the ferromagnetic signature as arising from the second bilayer.  By varying the magnetization of the second bilayer in our model, consistency with the spectral shape of the XRMS data of Fig.\ \ref{scxrmsvsth}(b) is only found for the full bulk magnetization in the second bilayer, with an error bar of 20$\%$.  
Additional evidence that the surface magnetization is barely distinguishable from the bulk comes from comparing the temperature dependence of the XMCD and XRMS intensities with that of the bulk magnetization.  Except for small differences within a few Kelvins of T$_C$, these are in perfect agreement (see Fig.\ \ref{xmcdxrmsvsT}).  This is in stark contrast to all other half-metallic oxides \cite{coeyrev} in which the loss of surface spin polarization is accompanied by significantly different temperature dependence.  For example, in the La$_{1-x}$Sr$_x$MnO$_3$ system, the Ônear surfaceÕ shows significant degradation from the bulk well below T$_C$ (see Sec.\ \ref{lsmosto-sec} below), and the actual surface displays a nearly linear decreasefor all T up to T$_C$ \cite{ref12} .  Such a strong temperature-dependent degradation of the spin polarization below that of the bulk magnetization is also found at interfaces in La$_{1-x}$Sr$_x$MnO$_3$ magnetic tunnel junctions, which show a complete loss of magnetoresistance for T less than $\sim$0.7 T$_C$ \cite{ref5,ref8,ref10,ref11,ref11_1}.  

\begin{figure}[h]
	\begin{tabular}{cc}
		\includegraphics[scale=.4]{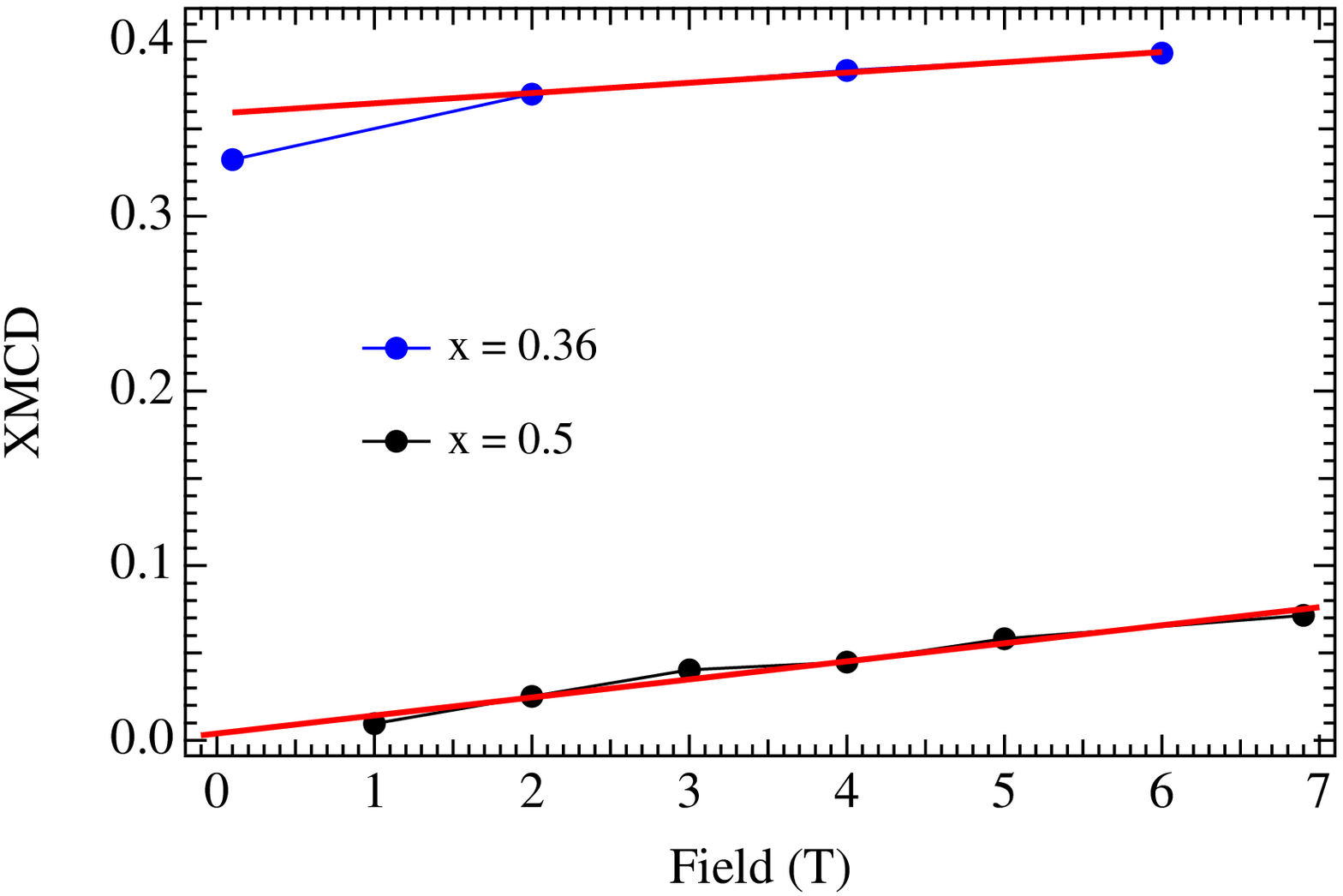} &
		\includegraphics[scale=.4]{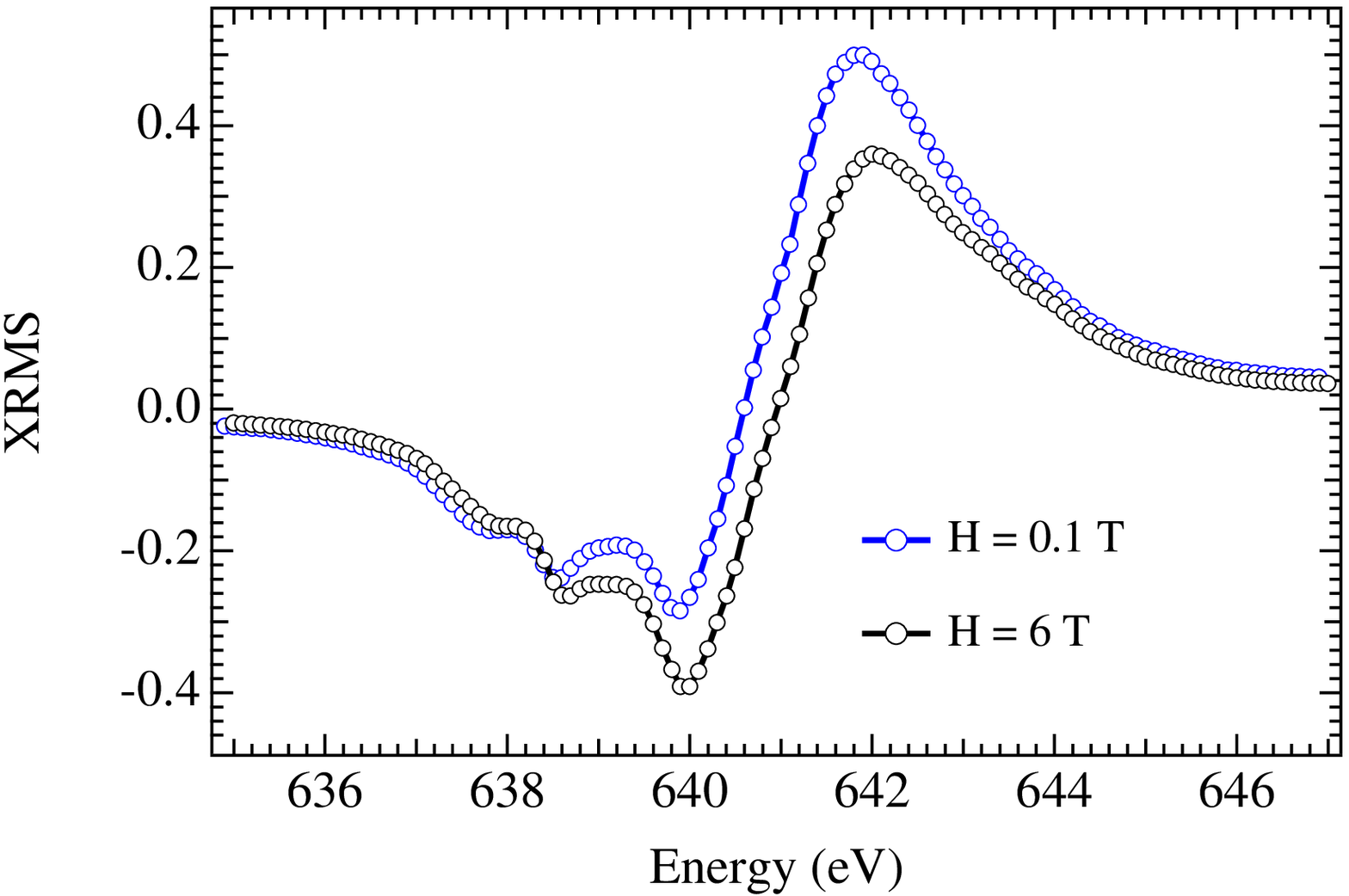} \\
		(a) &
		(b) \\
	\end{tabular}
\caption{(a) Near surface XMCD as a function of applied field for x=0.36 (ferromagnet) and x$\simeq$0.5 (A-type antiferromagnet at 4K). (b) The XRMS measured at T = 4K for 0.1 and 6 T. Note there is only a slight change in the lineshape indicating only a minor modification of the magnetization profile due to a small moment in the top layer from canting the AFM surface phase. }
\protect\label{xmcdvsh} 
\end{figure}

 Up to this point we have addressed the lack of long range FM order in the surface layer. However, due to the ever present t$_{2g}$ superexchange \cite{good}, antiferromagnetic (AFM) order at low temperatures is quite probable. Direct dectection of AFM order in a single surface layer is non-trivial, so we turned to high magnetic field measurements of the near surface XMCD to see if any canting could be induced. In Fig.\ \ref{xmcdvsh}(a), the near surface XMCD peak height is plotted as a function of applied field. There is a large change up to 2T, which bulk magnetometry of the x=0.36 shows is due to saturation of the bulk moment. However, the linear slope above 2T is not seen in bulk measurements and is attributed to canting of the anti-ferromagnetic phase in the surface layer.  It seems reasonable that the enhanced surface magnetization at 7 T results from overcoming the AF superexchange energy (typically ~10$^8$ erg/cm$^3$ in these materials \cite{tperring}) in the outermost bilayer.  To confirm this idea, we measured an insulating layered manganite with x $\simeq$ 0.5, which is an A-type AF below 200 K \cite{jfm2}. Here the A-type structure consists of two sheets of FM spins within the bilayer that are oriented antiferromagnetically. While there is no net moment (XMCD) at zero field, a magnetic field cants the FM sheets to produce a moment and the resulting surface susceptibility is linear with a value close to that of the high field component of the surface bilayer for x=0.36 (see Fig. \ \ref{xmcdvsh}(a)). As shown in Fig.\ \ref{xmcdvsh}(b), measurements of the XRMS at low and high fields shows only a slight change in the lineshape indicating that the magnetic profile is mostly unchanged. Modeling the data shows a slight net magnetization in the surface bilayer corresponding to the increase seen in the near surface XCMD due to canting the AF state. From this we conclude that the surface bilayer for x=0.36 exhibits AF order, presumably because the double exchange is preferentially weakened, allowing the superexchange to dominate.  

\begin{figure}[h]
\begin{tabular}{cc}
	\includegraphics[scale=.4]{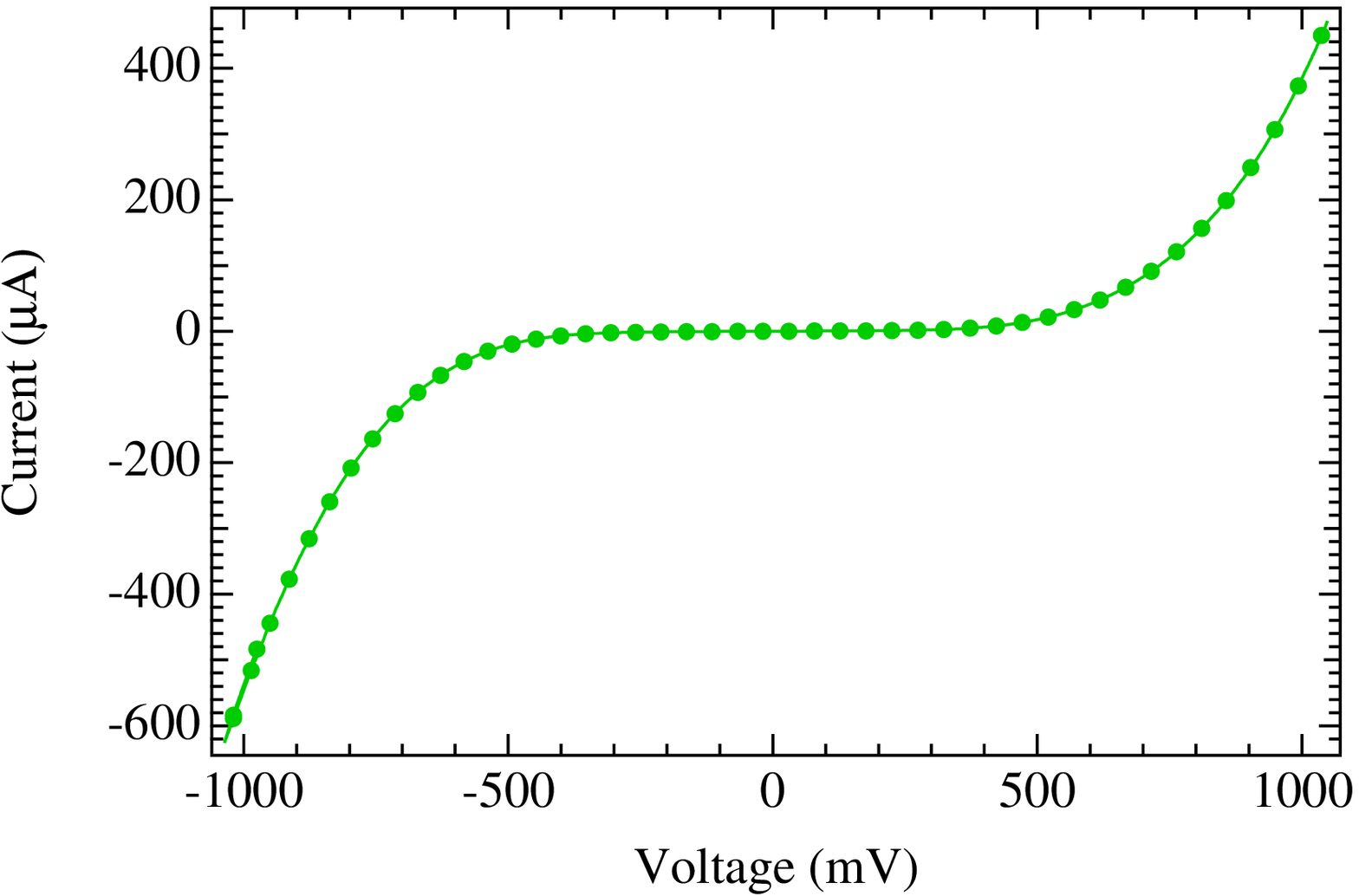} &
	\includegraphics[scale=.4]{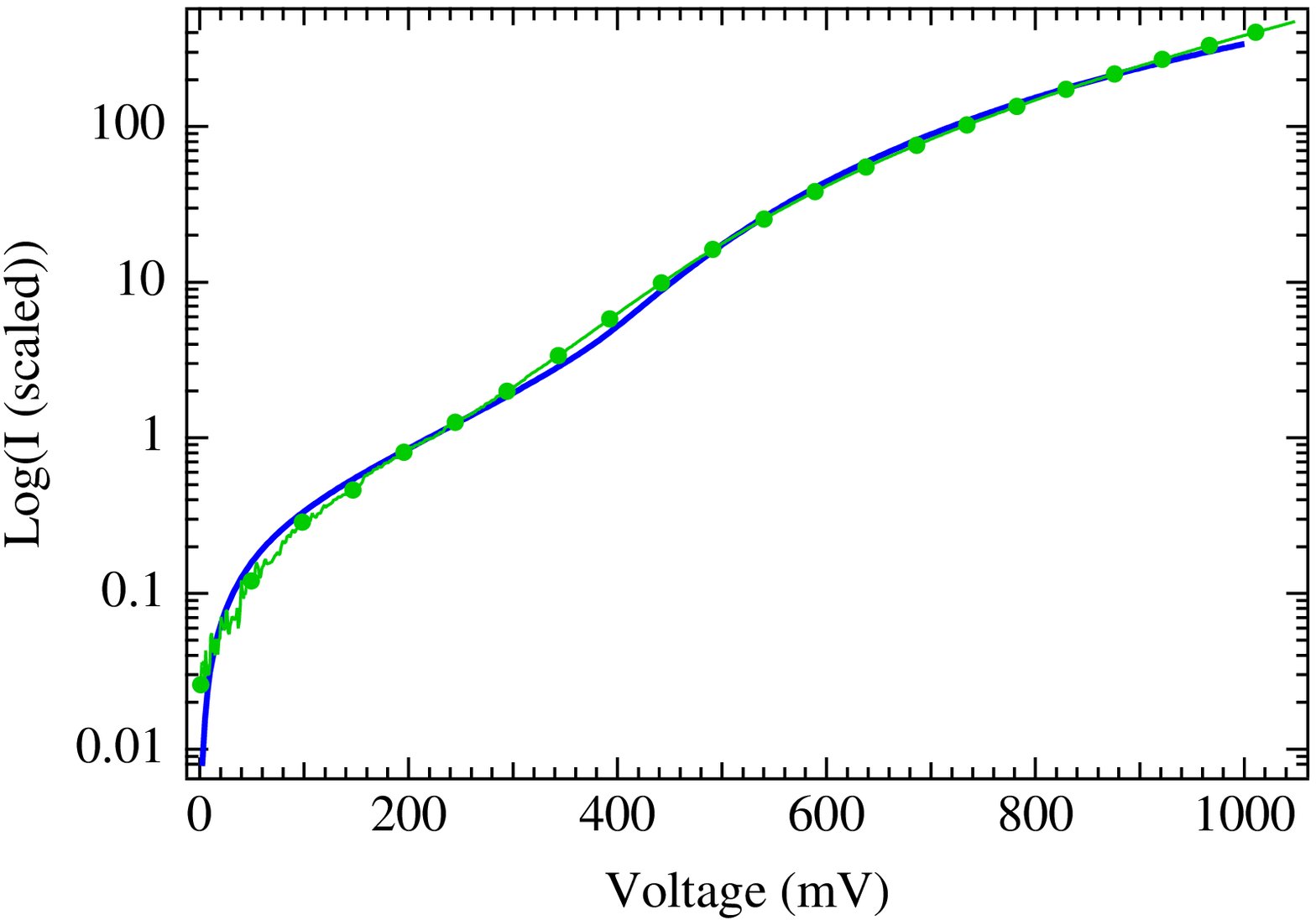} \\
	(a) &
	(b) \\
\end{tabular}
\caption{(a) Point contact tunneling I-V curve taken with a Au tip at 4K for the x=0.36 sample. This shows the large plateau due to the insulating surface bilayer. (b) Log-scale plot together with best fit to square barrier tunneling model. The fit parameters agree well with the surface bilayer being insulating as expected due to the lack of ferromagnetic order.}
\protect\label{pct} 
\end{figure}

From our understanding of the interplay between metallicity and ferromagnetism in the manganites \cite{zener,anderson}, we might expect that the non-FM surface bilayer is also non-metallic.  To assess the metallicity of the surface, we utilize point contact tunneling (PCT) with a Au tip.  At low temperature (4K), the measured I-V characteristic displays a prominent plateau at low voltage as shown in Fig.\ \ref{pct}(a). Given the magnetic result above showing lack of ferromagnetism in the surface bilayer, this plateau can be attributed to an insulating surface layer leading to a thin tunneling barrier. To confirm this idea, the data were modeled using a square barrier calculation, which showed very good agreement over several orders of magnitude (see Fig.\ \ref{pct}(b)). The resulting best fit parameters correspond to a barrier height of $\sim$350 meV and thickness of $\sim$1.4 nm.  Note that this thickness agrees well with the region of non-ferromagnetic order in the surface bilayer and is determined by a completely different approach. 

Therefore we conclude that the $\sim$1-nm-thick surface bilayer is an insulator.    The simultaneous absence of magnetic order and metallic conductivity in the top bilayer is consistent with the double-exchange mechanism used to explain conductivity below T$_C$ in the manganites \cite{zener,anderson}.  The abrupt changes in only the topmost bilayer are likely due to the weak electronic and magnetic coupling between bilayers that are engendered by the crystal structure and are absent in nonlayered magnetic oxides. In contrast to this, in the next section we will present the results for the 3D perovskite where the surface is not weakly connected to the bulk. 

\section{3D Perovskite Manganite (La$_{2/3}$Sr$_{1/3}$MnO$_3$)/SrTiO$_3$ Interface}
\label{lsmosto-sec}

As noted above, 3D manganites (e.g., La$_{2/3}$Sr$_{1/3}$MnO$_3$) have been extensively studied as possible half metallic contacts for use in spin based electronics. However, the general observation is that ferromagnetism is not stable at surfaces and interfaces \cite{ref5,ref8,ref10,ref11,ref11_1,ref12}. Here we seek to form a picture of the magnetic profile in order to better understand why and how magnetism is suppressed at these boundaries. 
In order to connect to the results seen in the magnetic tunnel junction work \cite{ref5,ref8,ref10,ref11,ref11_1} and to protect the surface from degradation, we decided to  cover the films with two unit cells of STO. 

\begin{figure}[h]
\begin{tabular}{cc}
	\includegraphics[scale=.4]{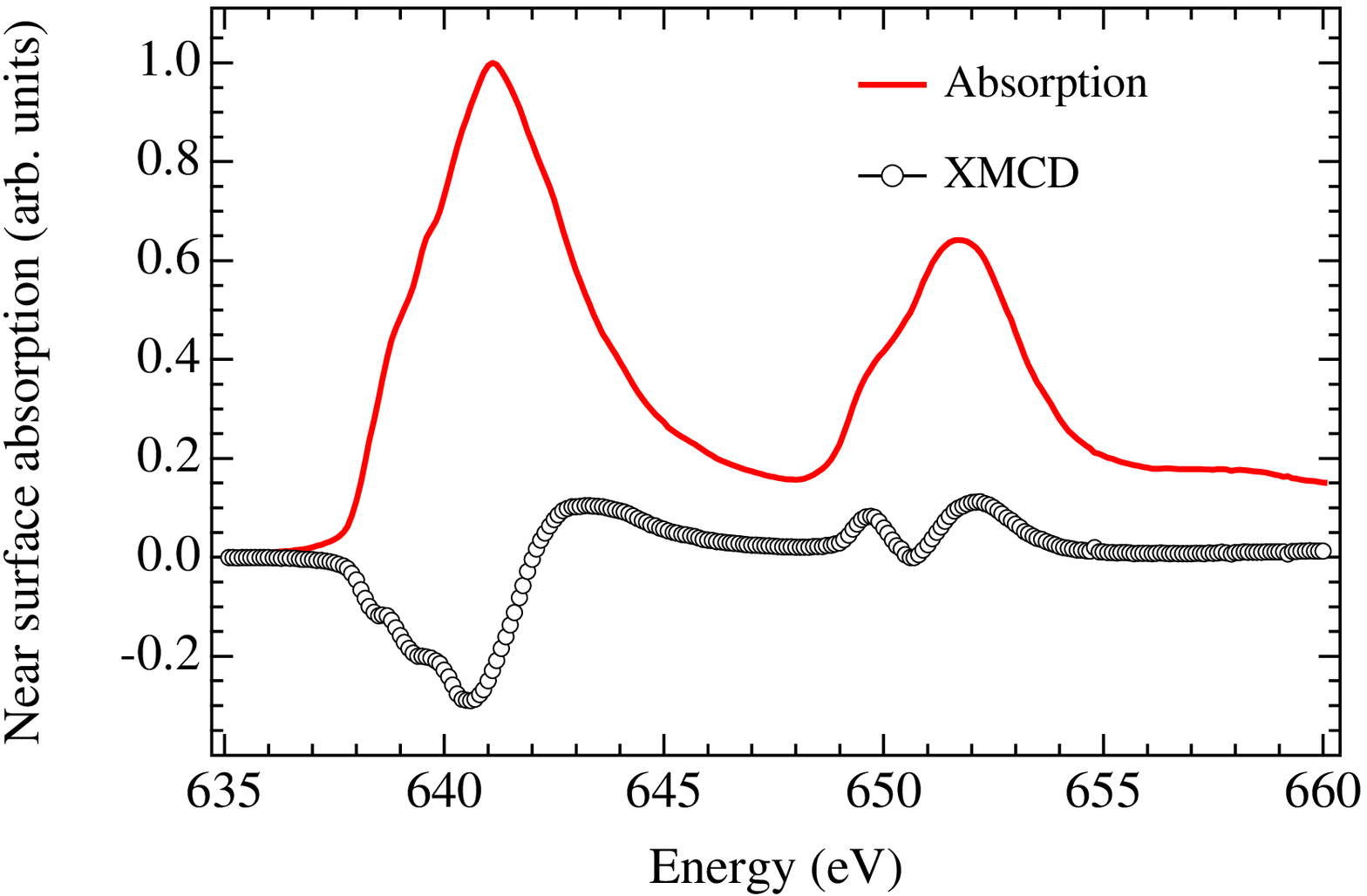} &
	\includegraphics[scale=.4]{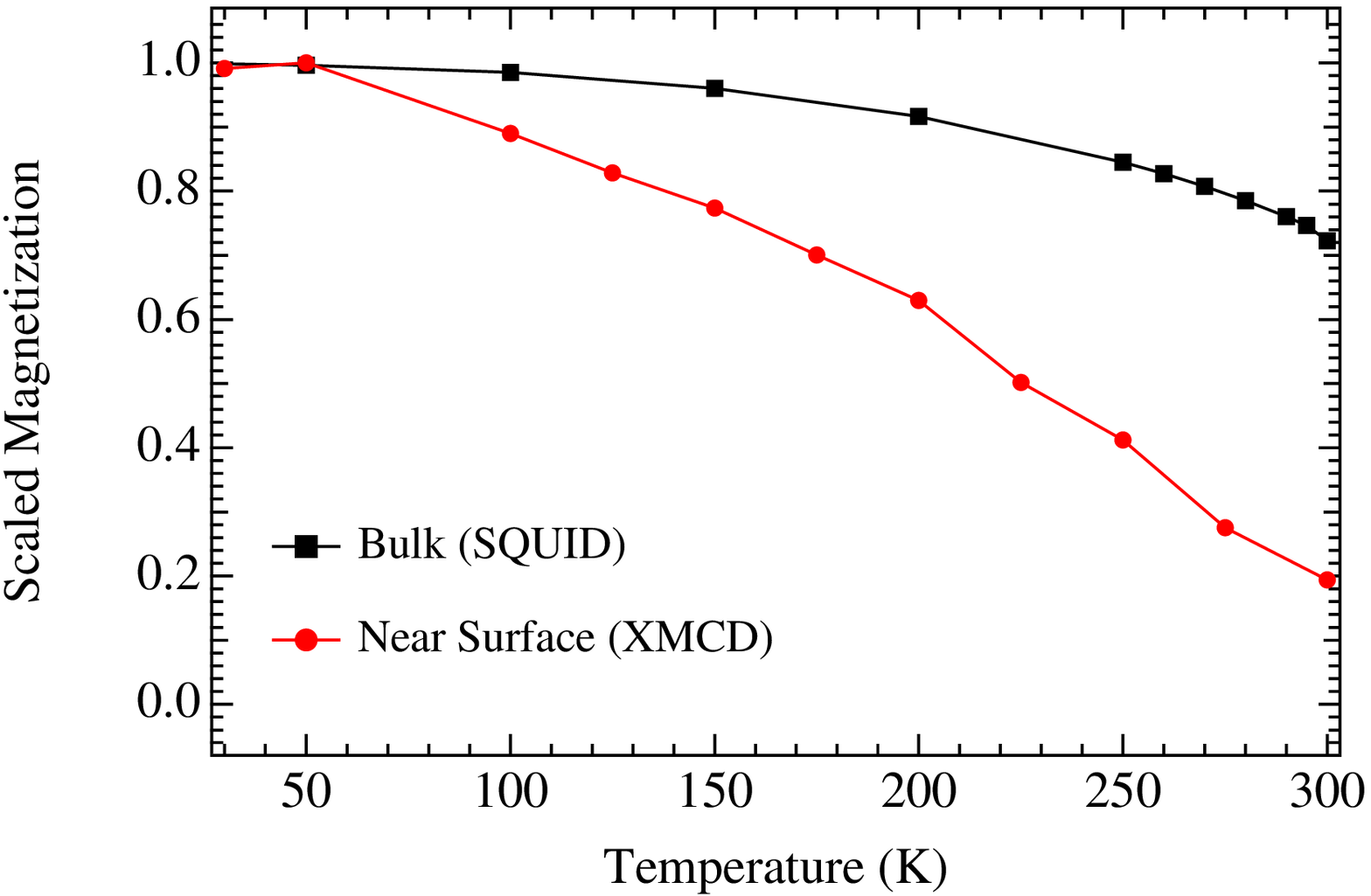} \\
	(a) &
	(b) \\
\end{tabular}
\caption{(a) Average absorption and XMCD data measured in the near surface mode. The good correspondence with data for the bilayer manganite and published data for the free LSMO surface \cite{ref12}, indicate a high quality LSMO/STO interface. (b) Temperature dependence of the near surface XMCD peak height and the bulk magnetization showing the strong suppression of the interface magnetism with increasing temperature.}
\protect\label{stoxmcddata} 
\end{figure}

First, we present the average behavior of the near surface region using measurements of the XMCD.  
The absorption data are quite consistent with data published on free surfaces prepared {\it in-situ} \cite{ref12} and indicate a high quality LSMO/STO interface (see Fig.\ \ref{stoxmcddata}(a)). Tracking the temperature dependence of the Mn L$_3$ XMCD peak height provides a means to monitor the magnetization near the interface. The results shown in Fig.\  \ref{stoxmcddata}(b) present clear evidence of a faster drop of the magnetization near the interface as compared to the bulk, which is consistent with the MTJ data \cite{ref5,ref8,ref10,ref11,ref11_1}  and of the same order as the effect observed for the case of a free surface \cite{ref12}. This implies that to first order, the LSMO/STO interface is similar to the case of a free LSMO surface, and the degraded magnetization may be due to the same underlying physics. However, these measurements are an average over several nm within the surface. To understand the problem in more detail, we need to map the depth-dependent magnetic profile to understand how the magnetization is suppressed as it approaches the surface.

\begin{figure}[h]
\begin{tabular}{cc}
	\includegraphics[scale=.4]{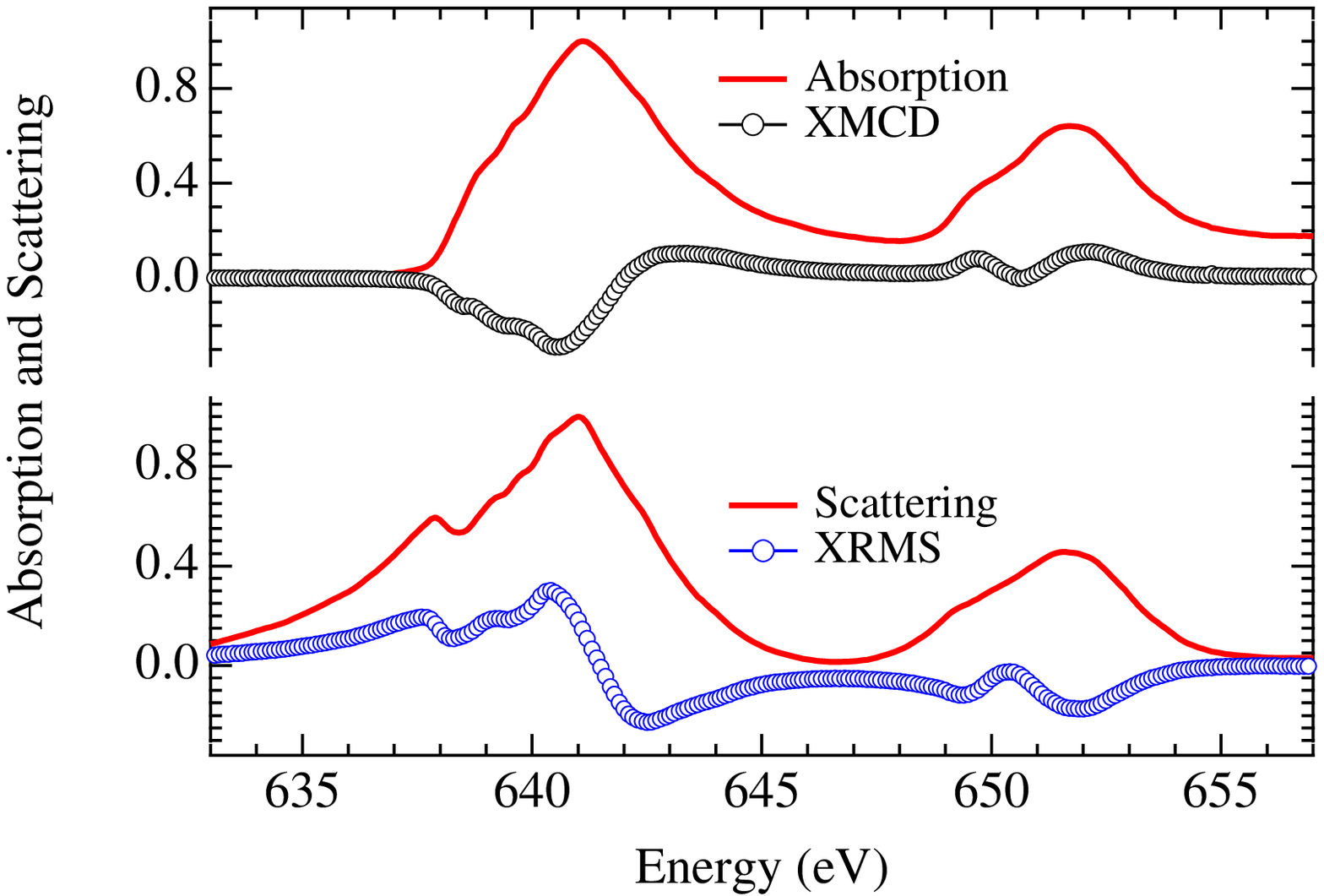} &
	\includegraphics[scale=.4]{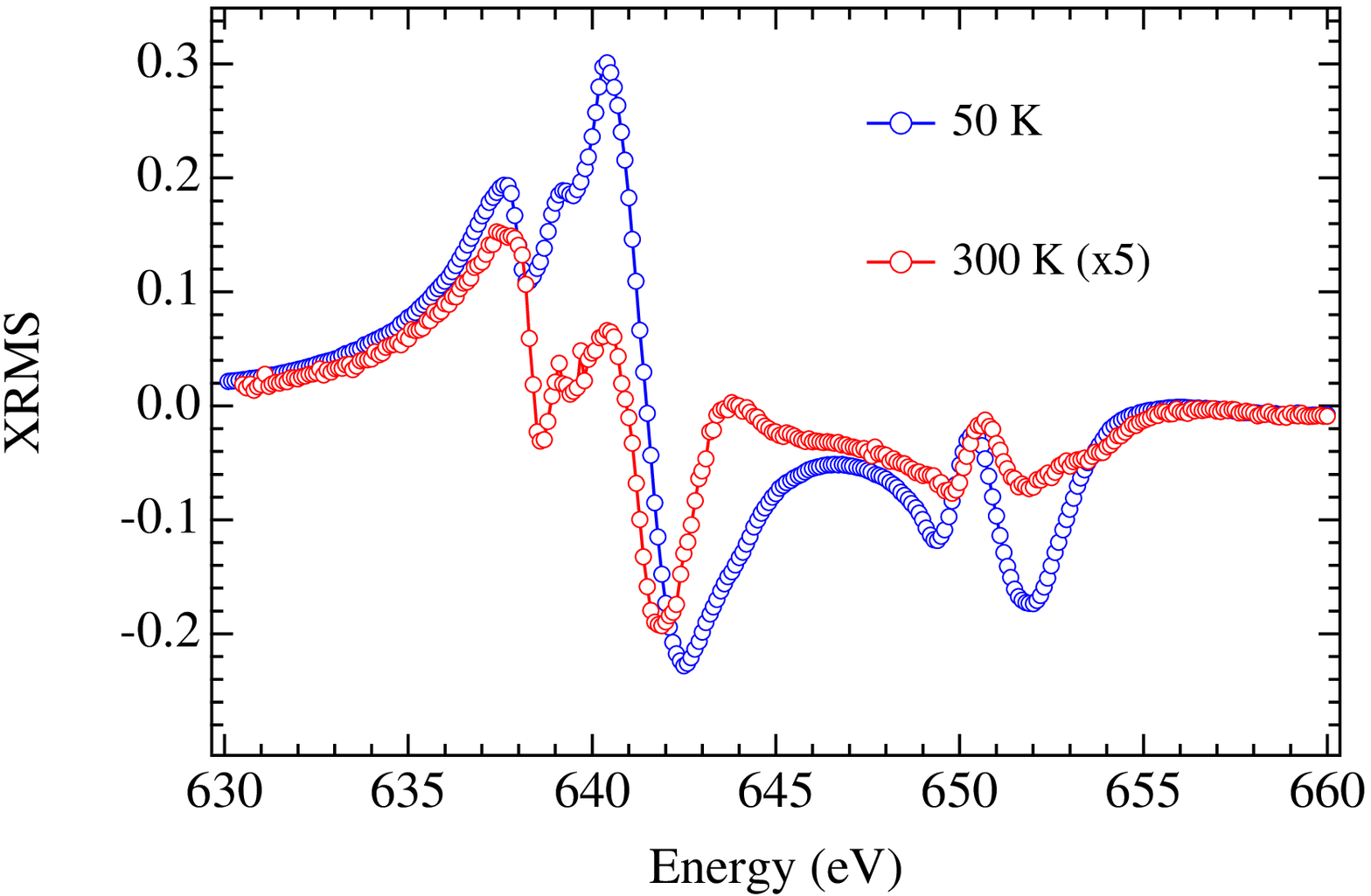} \\
	(a) &
	(b) \\
\end{tabular}
\caption{(a) Comparison of absorption and scattering data at $\theta$=11 deg.\ and T = 50 K. (b) XRMS data at 50 and 300 K showing the lineshape change that is attributed to the change in the magnetic profile with increasing temperature.}
\protect\label{stodata} 
\end{figure}

To construct the shape of the magnetization profile, we again utilize XRMS which probes the depth dependence of the magnetization. A comparison of the absorption and resonant scattering is shown in Fig.\ \ref{stodata}(a). As discussed above, the lineshape of the XRMS is dictated by two quantities: the scattering factors and the magnetic profile. Since the absorption lineshape, which is used to determine the scattering factors, does not change with temperature, we conclude the lineshape change shown in Fig.\ \ref{stodata} is due to a change in the magnetic profile at elevated temperatures.  The XRMS is modeled using a quasi-cubic perovskite unit cell that contains only one Mn ion, and the distance between possible magnetic scattering planes is fixed by the lattice constant ($\sim$0.4 nm).  Explanation of the XRMS depends on fitting the spectra by varying the average magnetization at each magnetic MnO$_2$ plane, which is specified by its depth in the sample with the STO interface being z=0.  If the incident angle and chemical boundaries are known, the detailed shape of the XRMS depends only on the shape of the magnetization profile.  

\begin{figure}[h]
\begin{tabular}{cc}
	\includegraphics[scale=.4]{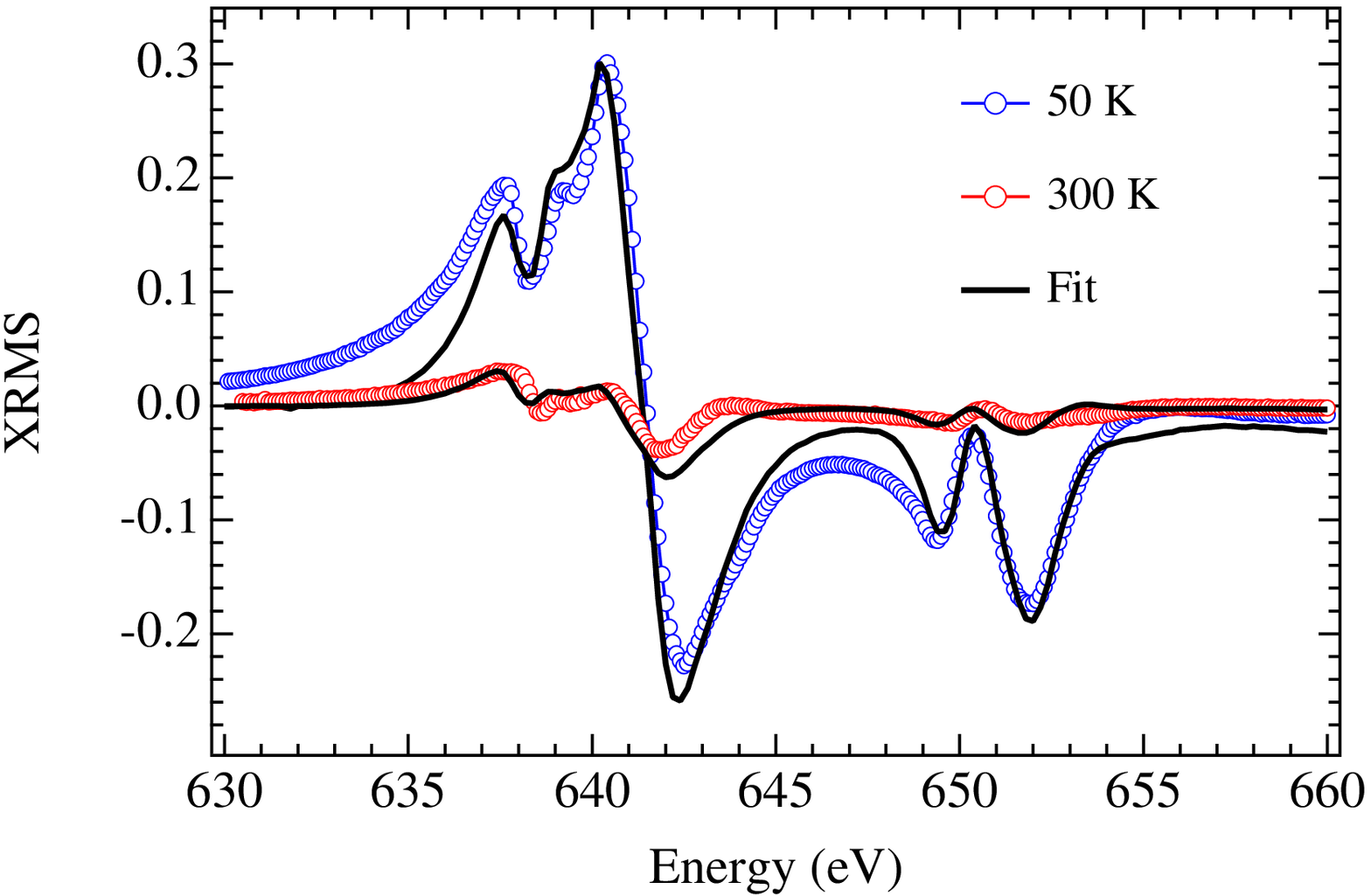} &
	\includegraphics[scale=.4]{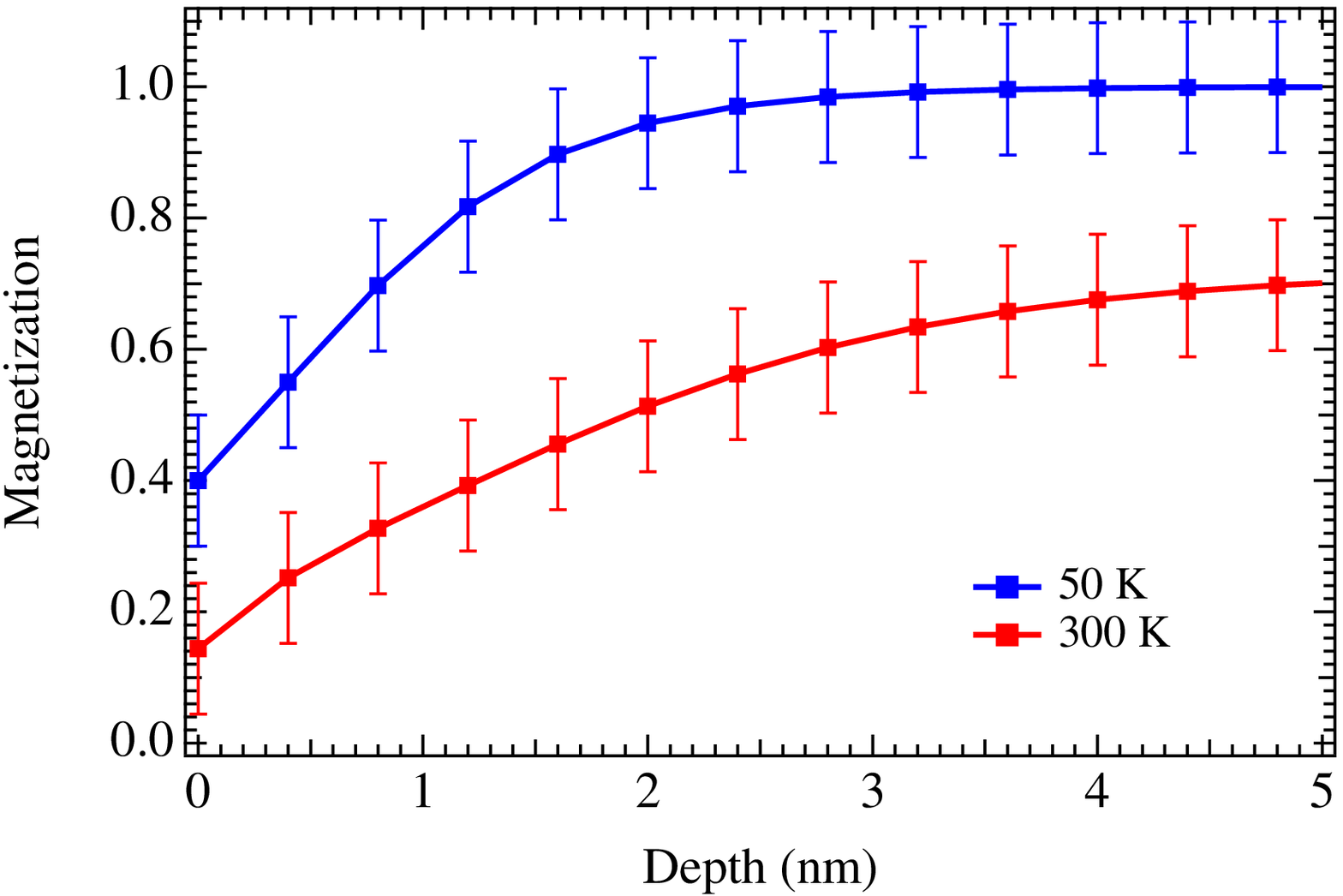} \\
\end{tabular}
\caption{(a) XRMS data and fits. (b) Magnetization profiles corresponding to the fits shown in (a).}
\protect\label{stofit} 
\end{figure}

After modeling many magnetic profiles with different functional forms, this methodology has had remarkable success in fitting the observed data (see Fig.\ \ref{stofit}(a)) and leads directly to the depth-dependent magnetic profile (see Fig.\ \ref{stofit}(b)). At low temperture, the profile demonstrates a continuous decrease from bulk ferromagnetism to $\sim$40$\%$ of the bulk value at the interface with a transition region of 1-2 nm ($\sim$3 u.c.). In the high temperature regime, the magnetization at the surface decays faster than the bulk magnetization. The profiles also demonstrate a reversible nature as a function of temperature, indicating there is a dynamic component to the loss of ferromagnetism near the interface. To check the model, we weighted the magnetic profile by the electron escape probability to reproduce the temperature-dependent results from the near surface average magnetization (see Fig.\ \ref{stoxmcddata}(b)), showing consistency between two independent measurements of the surface magnetism.

The reduction of ferromagnetism for the case of the La$_{2/3}$Sr$_{1/3}$MnO$_3$ surface can be understood by considering that the loss of coordination at the surface will result in reduced hopping, which will decrease the double exchange and disrupt the balance with the underlying AFM t$_g$ superexchange \cite{good}. This idea is supported by recent calculations of magnetic surface states of LSMO, which show a modified orbital occupancy near the surface and the lowest energy state corresponding to an insualting surface layer with magnetization AFM coupled to that of the bulk magnetization \cite{theory3,theory4}.  These results also find that the electronic structure returns to the bulk behavior within 3 u.c. of the surface, which agrees well with the lengthscale over which we see the suppressed magnetic order. However, in reality, due to disorder the surface region might result in an inhomogeneous lateral  structure at the interface. Since the XRMS profiles integrate over lateral variations, it is not possible with this measurement alone to determine the lateral variation of the magnetic order at the interface.  This measurement cannot distinguish between a uniformly reduced surface magnetization or an equal mixture of patches with full FM moment and AFM order, both of which result in the same laterally averaged magnetization. However, one can make the point that if the net moment were uniformly reduced, it would be difficult to retain the metallic state, which is clearly seen in the photoemission \cite{ref7,ref12} and tunneling results \cite{ref5,mbprl,ref8,ref10,ref11,ref11_1}. 

\section{Conclusions}

In conclusion, our results provide the general observation that ferromagnetism is not necessarily the lowest energy state at the surface or interface of manganites, which leads to a suppression or even loss of ferromagnetic order at the surface. This was illustrated with two cases ranging from the surface of the quasi-2D bilayer manganite (La$_{2-2x}$Sr$_{1+2x}$Mn$_2$O$_7$) to the 3D perovskite (La$_{2/3}$Sr$_{1/3}$MnO$_3$)/SrTiO$_3$ interface. For the bilayer manganite, which is ferromagnetic and conducting in the bulk, these probes present clear evidence for an intrinsic insulating non-ferromagnetic surface layer atop adjacent subsurface layers that display the full bulk spin polarization. This abrupt intrinsic magnetic interface is attributed to the weak inter-bilayer coupling native to these quasi-two-dimensional materials. This is in marked contrast to the non-layered manganite system (La$_{2/3}$Sr$_{1/3}$MnO$_3$/SrTiO$_3$), whose magnetization near the interface  is less than half the bulk value at low temperatures and decreases with increasing temperature at a faster rate than the bulk. 

\section{Acknowledgments}

Work at Argonne is supported by the U.S. Department of Energy, Office of Science, under Contract No. DE-AC02-06CH11357.

\end{document}